%% file: arxiv.tex
\documentclass[a4paper,USenglish,pdfa]{lipics-v2021}
\pdfoutput=1 
\hideLIPIcs  

\usepackage{float}
\usepackage{booktabs}
\usepackage[dvipsnames]{xcolor}
\usepackage{pgfplots}
\usepackage{pgfplotstable}
\pgfplotsset{compat=1.18}
\usepgfplotslibrary{groupplots}
\usepackage{tikz}
\usetikzlibrary{positioning,matrix,automata,angles,quotes,backgrounds,intersections,arrows,calc,tikzmark,shapes.geometric,arrows.meta}
\usepackage{xspace}
\usepackage[authoryear]{natbib}
\title{Hecate: A Modular Genomic Compressor}

\author{Kamila Szewczyk}{Algorithmic Bioinformatics, Saarland University, Saarbrücken, Germany \and Center for Bioinformatics, Saarland Informatics Campus, Germany \and \url{https://iczelia.net/}}{k@iczelia.net}{https://orcid.org/0000-0002-4002-9866}{}

\author{Sven Rahmann}{Algorithmic Bioinformatics, Saarland University, Saarbrücken, Germany \and Center for Bioinformatics, Saarland Informatics Campus, Germany \and \url{https://www.rahmannlab.de/people/rahmann}}{sven.rahmann@uni-saarland.de}{https://orcid.org/0000-0002-8536-6065}{}

\authorrunning{K. Szewczyk and S. Rahmann}  

\Copyright{Kamila Szewczyk and Sven Rahmann}  



\category{} 

\relatedversion{} 

\nolinenumbers  

\EventEditors{John Q. Open and Joan R. Access}
\EventNoEds{2}
\EventLongTitle{42nd Conference on Very Important Topics (CVIT 2016)}
\EventShortTitle{CVIT 2016}
\EventAcronym{CVIT}
\EventYear{2016}
\EventDate{December 24--27, 2016}
\EventLocation{Little Whinging, United Kingdom}
\EventLogo{}
\SeriesVolume{42}
\ArticleNo{23}

\newcommand{\hecate}{\textit{hecate}\xspace}

\begin{document}
\maketitle

\begin{abstract}
We present \hecate, a modular lossless genomic compression framework.
It is designed around uncommon but practical source-coding choices.
Unlike many single-method compressors, \hecate treats compression as a conditional coding problem over coupled FASTA/FASTQ streams (control, headers, nucleotides, case, quality, extras).
It uses per-stream codecs under a shared indexed block container.
Codecs include alphabet-aware packing with an explicit side channel for out-of-alphabet residues, an auxiliary-index Burrows-Wheeler pipeline with custom arithmetic coding, and a blockwise Markov mixture coder with explicit model-competition signaling.
This architecture yields high throughput, exact random-access slicing, and referential mode through streamwise binary differencing.

In a comprehensive benchmark suite, \hecate provides the best compression vs.\ speed trade-offs against state-of-the-art established tools (MFCompress, NAF, bzip3, AGC), with notably stronger behaviour on large genomes and high-similarity referential settings.
For the same compression ratio, \hecate is 2 to 10 times faster.
When given the same time budget as other algorithms, \hecate achieves up to 5\% to 10\% better compression.
\end{abstract}

\newcommand{\hadd}{0.3}
\input{content}
\bibliography{reference}

\appendix

\end{document}

%% file: content.tex

\section{Introduction}

The year 1990 saw the launch of the International Human Genome Project, foreshadowing the start of the era of high-throughput sequencing.
Three years after, the first specialised genomic compression tool, \textit{Biocompress} \citep{grumbach1993}, was developed.
Since then, the Sequence Read Archive (SRA) observed the acquisition of 12 petabases of new public data in the year 2024 \citep{shiryev2024}.
In 2026, the persistent systems gap has become apparent:
Sequencing throughput has accelerated faster than practical storage and transport infrastructure.

This gap is not explained solely by missing algorithms.
Many researchers and practitioners still rely on universal compression methods (\textit{pigz} \citep{adler2023} or \textit{lbzip2} \citep{izdebski2015} provide parallel implementations of popular \textit{gzip} and \textit{bzip2} codecs) or even uncompressed formats.
In many production pipelines, the bottleneck is primarily architectural: tools are either monolithic (one model class everywhere) or operationally awkward (weak random access, brittle indexing, limited integration).

C.~G. Nevill-Manning and I.~H. Witten demonstrated in their 1999 paper \textit{"Protein is incompressible"} \citep{755675} that increasing context (i.e., Markov model order) for predicting the next amino acid yields sharply diminishing returns.
Their result does not preclude advancements in one-shot single genomic sequence compression, as they frequently contain substantial amounts of non-coding regions.
The paper does however place rigid and disappointing bounds on the effectiveness of reference-free techniques.
This result shifts our focus towards referential compression.
Tools such as HRCM \citep{Yao2019} or AGC \citep{deorowicz2023agc} have demonstrated impressive compression ratios on highly similar genomic sequences, such as large collections of individual human genomes.
According to \citet{auton2015global}, the typical difference between an individual's genome and a reference human genome was estimated at around 20~million base pairs, or 0.6\% of the total size.
Consequently, \citet{christley2009human} provides an amusing example of referential compression that, under the assumption that the recipient has access to an agreed-upon reference genome, allows for transmitting a person's genome in the size of an e-mail attachment.

We briefly position \hecate against notable genomic compressors.
DELIMINATE \citep{mohammed2012deliminate} uses two-symbol position coding followed by external compression.
MFCompress \citep{pinho2014mfcompress} and related Markov tools focus on fixed-order context models.
NAF \citep{kirill2019nucleotide} prioritizes fast decoding through fixed 4-bit packing plus zstd.
JARVIS3 \citep{Sousa2024} (linear mixing of repeat and Markov models to provide probabilities to an arithmetic coder) and GeCo3 \citep{Silva2020} (neural network mixing of \{direct, hashed, substitution-tolerant\}-Markov models to compress genomic sequences) employ richer probabilistic mixtures.
AGC~\citep{deorowicz2023agc} is strong for assembled-genome collections with fast queries.

Our goal differs in scope:
We engineer one modular genomic container framework that supports both high-performance reference-free operation and also a strong referential mode.
Moreover, we focus on flexibility on the systems level with stream-level method assignment, index tiers, block-local validation, and efficient slicing.

As a result, on typical FASTQ files, which mix different streams of information, such as headers, sequences and quality information, \hecate obtains either smaller compression ratios or faster compression speeds than existing tools.

In more detail, our main contributions are as follows.
First, as mentioned, we introduce a modular genomic container with stream-factorized coding, indexed random access, and integrity semantics aligned with block independence.
Second, we provide a new highly engineered codec (\textit{hecate-bwt}) based on the Burrows-Wheeler transform (BWT).
It is a large-block BWT pipeline with auxiliary indices, dynamic 32/64-bit suffix array paths, and a sophisticated post-BWT encoding stage.
Third, we introduce another engineered codec (\textit{markov-mix}) based on Markov mixtures and arithmetic coding. It is an explicit blockwise model-competition codec with heterogeneous update asymmetry and reverse-context coupling in deep models.
Fourth, we contribute a reference-based compression mode for medium to large genomes that performs streamwise binary differencing separately over semantic channels, preserving codec modularity after patch reconstruction.

\section{Background}
\label{sec:background}

\subsection{Burrows-Wheeler Transform and LF Mapping}

Given a block $T[0..n-1]$, let $SA$ be the suffix array of $T$.
The BWT output $L$ is defined by
\[
L[i] = T[(SA[i]-1) \bmod n],
\]
which is equivalently the last column of the lexicographically sorted rotation table.
Two equivalent conventions are common: either append a unique end marker (so the original row is identified by that marker), or omit an explicit marker and store a primary index $p$ such that $SA[p]=0$.
The transform is reversible and tends to create locally run-rich output, which is easier to entropy-code well.

Inversion is driven by the LF mapping.
Let $C[a]$ be the number of symbols in $L$ that are lexicographically smaller than symbol $a$, and let $\mathrm{Occ}(a,i)$ be the number of occurrences of $a$ in $L[0..i-1]$.
Then
\[
\mathrm{LF}(i) = C[L[i]] + \mathrm{Occ}(L[i], i).
\]
Starting from the identified original row (via marker or primary index) and iterating LF reconstructs the original text in reverse order.
This LF machinery is the core of the FM index \citep{892127}.
In practice, long blocks can additionally store sparse restart anchors so inversion does not need to follow LF from a single start point over long distances.

\subsection{Prediction by Partial Matching (PPM)}

Prediction by Partial Matching (PPM) \citep{1096090} models the next symbol using a hierarchy of context lengths.
For each position, it first queries the longest available context (highest order).
If the symbol has not been observed there, the encoder emits an escape and backs off to shorter contexts until a distribution is found (or a lowest-order base model is reached).

The selected distribution is then encoded with an entropy coder, typically arithmetic or range coding.
This per-symbol backoff makes PPM robust on heterogeneous data, but frequent escapes add coding overhead and can weaken very deep sparse contexts.

\section{Methods: Reference-Free Compression}

The \hecate framework provides three main codec families for reference-free compression, chosen to cover distinct regions of the ratio/speed frontier:
Lempel-Ziv (via \textit{zstandard} \citep{zstd}), Burrows-Wheeler Transform (\textit{hecate-bwt}), and blockwise Markov mixture coding (\textit{markov-mix}).
The rationale follows well-established empirical knowledge \citep{mahoney2005large}:
Lempel-Ziv methods decompress fast at the expense of ratio, BWT-based methods offer a strong middle ground, and statistical models push ratio further at higher computational cost.
Each codec operates on semantically factored streams under a shared container, so the choice of backend can be made per-stream without changing the file format.

\subsection{Container and Preprocessing}

At encode time, \hecate factors a FASTA/FASTQ input into semantic streams: control (\texttt{CTRL}), headers (\texttt{HDR}), nucleotide payload (\texttt{NUC}), case (\texttt{CASE}), quality (\texttt{QUALITY}), and an extra (non-IUPAC) channel (\texttt{EXTRA}).
The container then allows for assignment of each stream to a codec independently: We suggest BWT or Markov models for nucleotides, zstd or raw coding for low-entropy control and header streams, and dedicated coders (for example the built-in \texttt{rans-o1}) for quality.
This factorization reduces the modeling problem from one heterogeneous joint process into several narrower conditional processes with codec-specific inductive bias.

As a preprocessing step, the nucleotide stream is packed into a 2-bit or 4-bit representation, yielding a 75\% or 50\% size reduction before any entropy coding.
Out-of-alphabet characters are handled exactly through the \texttt{EXTRA} side channel:
Each non-IUPAC character emits a delta-coded position and one extra byte.
For packing width $k \in \{2, 4\}$ bits, this is beneficial whenever the fraction of non-ACGT symbols satisfies $e/n < (8-k)/40$.
Concretely, less than 15\% for 2-bit and less than 10\% for 4-bit packing.
Real genomic data falls well below these thresholds.
The packing stage runs at approximately 500\,MB/s in encoding and 3--10\,GB/s in decoding (single core, depending on output format and memory-mapping availability), so it never dominates end-to-end cost.

\subsection{hecate-bwt}
\label{sec:hecate-bwt}

The \textit{hecate-bwt} codec is a BWT-based codec designed to overcome specific limitations of existing block-sorting compressors.
The older and well-established \textit{bzip2} uses small blocks (900\,KB) paired with a relatively weak post-BWT stage (imprecise symbol ranking, run-length coding, Huffman coding), which limits compression ratio on large genomes.
The newer \textit{bzip3} \citep{szewczyk2022b} supports larger blocks (up to 512\,MB) and a better context model (partially re-used in this work, originating from Ilya Murvayev's \textit{bcm} algorithm), but carries always-enabled filtering stages that hurt genomic compression, limited parallelism within BWT construction, slow serial decoding, and high memory usage.
The \textit{bsc} compressor \citep{grebnov2009} achieves strong ratios via a fast post-BWT coder, but is considerably more complex and typically compresses slightly worse than \textit{bzip3}.
Finally, the \textit{bbb} compressor \citep{bbb} resolves the small-block and weak-coder limitations of \textit{bzip2}, but at the cost of very slow encoding and decoding.

\paragraph{BWT construction and auxiliary-index inversion.}
\textit{hecate-bwt} supports block sizes practically limited only by available memory.
BWT construction and inversion use the parallel suffix sorting algorithm \textit{libsais} \citep{libsais}, which dispatches to 32-bit or 64-bit suffix array paths from the actual block size.
For small blocks ($n < 32$\,KiB), a classic primary index $p$ suffices for inversion.
For larger blocks, we use auxiliary-index BWT with stride
\[
r = 2^{\left\lfloor \log_2 \max\!\left(1, \lfloor n/8 \rfloor\right) \right\rfloor},
\]
and serialize a fixed 256-entry auxiliary table $A$ \citep{OHLEBUSCH201421}.
These entries act as deterministic restart scaffolds during inversion: rather than following the LF mapping from a single primary index $p$, inversion can resume from precomputed anchors, bounding the restart distance by $\mathcal O(r)$.
This gives practical inversion locality while keeping the auxiliary payload fixed at 256 words regardless of block size.

For a block of $n$ bytes at SA width $w \in \{32, 64\}$ bits, the structural metadata overhead is
\[
L_{\mathrm{meta}}(n, w) = 8 + 16 + 256w + 64 + 64\left\lceil n/C \right\rceil \quad \text{bits},
\]
where $C = 2^{24}$ is the chunk size for independent arithmetic-coded chunks.
The overhead rate $L_\mathrm{meta}(n,w)/n$ is negligible for practical block sizes.

\paragraph{Bitwise probability model.}
BWT-based compressors differ considerably in performance depending on how they encode the result of the BWT. We present an encoding based on the arithmetic coding algorithm tailored to genomic data.

Each BWT output byte is coded MSB-first through a depth-8 binary context tree with a leading one bit (contexts $c \in \{1, \ldots, 256\}$). We perform this through the use of a predictor that estimates the probability of the next bit being 1 given the context.

Let $u^{(0)}_c$, $u^{(1)}_{p,c}$, $u^{(2)}_{q,c} \in [0, 65535]$ be the order-0 and two order-1 counters (with $p$ being the previous byte and $q$ the byte before that). Then, let $t$ be the current bit-wise position in the BWT, and consequently $b_t \in \{0,1\}$, $c_t$, $p_t$ and $q_t$ be the current bit, context, previous byte and byte before that, respectively.
The base predictor blends these as
\[
\hat\Pr_t = \frac{6\bigl(u^{(0)}_{c_t} + u^{(1)}_{p_t,c_t}\bigr) + 4\,u^{(2)}_{q_t,c_t}}{16}.
\]
This is a fixed-weight mixture with effective contribution ratios $3{:}3{:}2$ for the three counter families, reflecting that direct and one-symbol-back contexts carry roughly equal predictive weight on BWT output, while the gapped context $u^{(2)}$ (which skips the directly preceding byte and shares its table with $u^{(1)}$) contributes a stabilizing third opinion.

Run state is modeled explicitly: $f_t = \mathbf{1}[\text{run length} > 2]$.
A run-conditioned secondary symbol estimation (SSE) table $s_{(2c_t + f_t), j}$ with $j \in \{0, \ldots, 16\}$ is linearly interpolated at the quantized base prediction:
\[
j_t = \left\lfloor \hat\Pr_t / 2^{12} \right\rfloor, \qquad
\lambda_t = (\hat\Pr_t \bmod 2^{12}) / 2^{12},
\]
\[
\hat{s}_t = (1 - \lambda_t)\, s_{2c_t+f_t,\, j_t} + \lambda_t\, s_{2c_t+f_t,\, j_t+1}.
\]
The final arithmetic split probability is $q_t = (\hat\Pr_t + \hat{s}_t) / 2^{17}$. Counters are initialized at $2^{15}$ (unbiased midpoint); SSE entries are seeded as an ordered grid $s_{c,j} \approx j \cdot 2^{12}$.

\paragraph{Counter dynamics.}
The three counter families use asymmetric exponential-move updates:
\[
U_\tau(v, b) =
\begin{cases}
v - \lfloor v / 2^\tau \rfloor, & b = 0, \\
v + \lfloor (65535 - v) / 2^\tau \rfloor, & b = 1.
\end{cases}
\]
We set $\tau = 3$ for $u^{(0)}$, $\tau = 5$ for $u^{(1)}$, and $\tau = 7$ for SSE entries. Each family thus acts as an exponential moving estimator with half-life $t_{1/2}(\tau) \approx (\ln 2)\, 2^\tau$: roughly 6, 22, and 89 symbols respectively. This three-rate separation is deliberate. Under long BWT runs (characteristic of repetitive genomic regions), the fast lane $u^{(0)}$ tracks local statistics, the medium lane $u^{(1)}$ provides context-conditioned stability, and the slow SSE lane resists transient fluctuations. The blend therefore approximates a multiscale estimator without explicit mixture weights.

The choice of fixed-step exponential moving average is dictated by both performance and the specifics of the Burrows-Wheeler Transform output distribution. A practical alternative would be the Krichevsky-Trofimov estimator~\citep{1056331}, which has a strong theoretical pedigree as a minimax optimal adaptive estimator for Bernoulli processes. However, its implementation issues a costly \texttt{idiv} instruction per update, which is prohibitive at the billions-of-symbol scale of BWT output. The fixed-step update, by contrast, can be implemented with fast bit shifts and additions with better cache behaviours ($512\times$ \texttt{u16} vs $512\times 2\times$ \texttt{u32}).

Simple and well known quantitative analysis for stationary sources shows that the effective sample size of the fixed-step estimator is $n_{\mathrm{eff}} \sim 2^\tau$ and $\text{Var}(p_\infty) = \frac{2^{-\tau}}{2-2^{-\tau}}\,\theta(1-\theta) = \frac{\theta(1-\theta)}{2^{\tau+1}-1}$. Notably, the variance is not correlated with the total amount of samples. Since variance does not vanish, excess log-loss remains and the total regret grows linearly in data size. This is juxtaposed with the Krichevsky-Trofimov estimators that minimise worst-case regret, converge with $\mathbb E[\hat \theta] \to \theta$ and $\mathbb E[\hat \theta] - \theta = \frac{1-2\theta}{N+2}$ (thus bias decreasing $\sim \mathcal O(1/N)$), and minimise per-symbol redundancy to $\mathcal O(\log N / N)$.

That said, the BWT output distribution is far from stationary. Even if the symbol streams have a strong local regularity, the predictors receive a binarized view of the data that contains many strong, structured non-stationarities: the distribution of the next bit heavily depends on whether we are in a run, the prefix and the current magnitude bucket. Further, probabilities can swing quickly and dramatically, especially at run boundaries. This motivates the use of ensembles of fixed-step EMA (sometimes called Vitter counters in the context of statistical data compression) with different $\tau$ parameters, which can be seen as a simple and efficient way to approximate a multiscale estimator that can adapt to both fast and slow distributional changes without the overhead of a full Krichevsky-Trofimov update.

\paragraph{Parallel chunked coding.}
The BWT output is partitioned into $\lceil n / C \rceil$ independent chunks of size $C = 2^{24}$ bytes. Arithmetic coding across chunks is embarrassingly parallel once the BWT is produced: chunk headers record per-chunk compressed sizes, enabling direct random-access decoding. Since $\lceil n / C \rceil$ is small for practical blocks, scheduling overhead stays subdominant and parallel efficiency is controlled primarily by chunk granularity and memory locality.

\subsection{markov-mix}
\label{sec:markov-mix}

\textit{markov-mix} is a statistical codec that treats each block of nucleotides as a local expert-selection problem. In contrast to per-symbol PPM backoff (Section~\ref{sec:background}), \textit{markov-mix} selects the best-performing expert per block of 80 symbols, then codes the expert index explicitly. Tables~\ref{fig:ppm_vs_markovmix} and~\ref{fig:mfc_vs_markovmix} summarize the structural differences.

\begin{table*}[t]
  \centering
  \small
  \caption{Feature comparison: PPM vs.\ markov-mix.}
  \begin{tabular}{@{}p{0.18\textwidth}p{0.37\textwidth}p{0.37\textwidth}@{}}
    \toprule
    Feature & PPM & markov-mix \\
    \midrule
    Models & $n$ models with fixed orders $0$ to $n-1$ & Curated set of heterogeneous orders and storage modes \\
    Model selection & Per-symbol basis using context tree and escape mechanism & Per-block basis using estimated coding cost \\
    Model updates & All models updated after each symbol & Winner receives full update; others receive asymmetric updates \\
    Data structure & Trie or suffix array with pointer-heavy traversal & Flat arrays with branch-light, SIMD-friendly updates \\
    High-order decisivity & Escapes dampen deep-order confidence & Deep context chosen when locally optimal; no escape overhead \\
    \bottomrule
  \end{tabular}
  \label{fig:ppm_vs_markovmix}
\end{table*}

\begin{table*}[t]
  \centering
  \small
  \caption{Feature comparison: MFCompress vs.\ markov-mix.}
  \begin{tabular}{@{}p{0.18\textwidth}p{0.37\textwidth}p{0.37\textwidth}@{}}
    \toprule
    Feature & MFCompress & markov-mix \\
    \midrule
    Estimator & $(\alpha + \beta \cdot \#(s, c))/(|\Sigma| \cdot \#(c))$; floating-point & $2^{\gamma} \cdot \#(s, c)/(|\Sigma| \cdot \#(c))$; fixed-point \\
    Model selection & Order-3 model with $\alpha = 1$, $\beta = 50$ & Order-5 model with $\gamma = 0$ \\
    Coding backend & Arithmetic coding (CACM with rescaling) & Range coding \\
    Context handling & Re-scanning of past data given a pointer & In-register, branchless rolling context updates \\
    IUPAC/unknown path & Interleaved in main stream & Binary unknown gate + dedicated unknown-symbol model \\
    Processing & Parallel (multi-file) & Serial (multi-core via block-level parallelism) \\
    \bottomrule
  \end{tabular}
  \label{fig:mfc_vs_markovmix}
\end{table*}

\paragraph{Expert family.}
The codec maintains five experts $\mathcal{M} = \{m_0, \ldots, m_4\}$, each parameterized by a tuple $(k, \alpha, \rho, \rho^{\mathrm{rc}}, c_{\max})$:
\[
\begin{aligned}
&(3, 0, 0, 0, 65535),\quad (7, 0, 0, 0, 1023), \\
&(11, 2, 0, 1, 255),\quad (15, 6, 1, 1, 15),\quad (13, 9, 1, 0, 0).
\end{aligned}
\]
Here $k$ is the context order, $\alpha$ controls count scaling, $\rho$ controls whether non-selected experts perform full count updates ($\rho=1$) or context-only updates ($\rho=0$), $\rho^{\mathrm{rc}}$ enables reverse-complement coupling, and $c_{\max}$ is the counter saturation threshold. Storage mode depends on context-space size: 16-bit counters for small spaces ($k = 3, 7$), 8-bit for medium/deep spaces ($k = 11, 13$), and packed 4-bit nibbles for very deep order ($k = 15$). This yields a total model state of approximately 2.45\,GB, dominated by the order-15 table at 2\,GB.

\paragraph{Per-context prediction.}
For expert $m$, context $h_t$, and symbol $s \in \{0,1,2,3\}$, counts $c_{m,h_t,s}$ induce frequencies
\[
f_{m,h_t,s} = 1 + 2^{\alpha_m} c_{m,h_t,s}, \qquad
P_m(s \mid h_t) = \frac{f_{m,h_t,s}}{\sum_{a=0}^{3} f_{m,h_t,a}}.
\]
The count scaling factor $2^{\alpha_m}$ amplifies the influence of observed counts in deep models, where context occupancy is sparse but observations are highly informative. Forward contexts evolve as a base-4 shift register: $h_{t+1} = ((h_t \bmod 4^{k-1}) \ll 2) + x_t$. When reverse-complement coupling is enabled ($\rho^{\mathrm{rc}} = 1$), a paired reverse context is updated simultaneously, which helps on strand-symmetric motifs.

\paragraph{Expert selection.}
Each expert is scored on a block $B = (x_1, \ldots, x_{80})$ using a log-table surrogate of cross-entropy:
\[
\tilde{C}_m(B) = \sum_{t=1}^{80} \bigl[\ln F_{m,h_t} - \ln f_{m,h_t, x_t}\bigr],
\]
computed in scaled fixed-point nats ($\times 10^6$). The chosen expert is $m^* = \arg\min_{m \in \mathcal{M}} \tilde{C}_m(B)$, and $m^*$ itself is range-coded by an order-5 auxiliary model over previous expert choices. In the worst case, the selector contributes $\log_2 5 \approx 2.32$ bits per block, but empirical selector entropy is much lower due to local persistence of expert identity -- genomic neighborhoods tend to stay within a single compositional regime across many consecutive blocks.

\paragraph{Asymmetric updates.}
The selected expert receives full symbol-count updates and context advance. Non-selected experts receive either context-only updates ($\rho=0$) or full updates ($\rho=1$), with reverse-complement context handling controlled by $\rho^{\mathrm{rc}}$. This asymmetry is a core design choice: deep experts maintain phase with local context drift even when not selected, while selective count updates control overhead. On counter saturation ($c_{m,h,s} = c_{\max} > 0$), all four symbol counters in that context are halved before incrementing the winning symbol; for $c_{\max}=0$, counters are reset before increment.

\paragraph{4-bit unknown-path factorization.}
In 4-bit mode, the codec explicitly factorizes the coding distribution for each block as
\[
\begin{aligned}
P(B) = P(z_B)\,P(m^*_B) \prod_{t=1}^{80} P(u_t)\,&P(x_t \mid u_t{=}0, m^*_B)^{\mathbf{1}[u_t=0]} \\
&\cdot P(v_t \mid u_t{=}1)^{\mathbf{1}[u_t=1]},
\end{aligned}
\]
where $z_B$ is a block-level flag indicating the presence of non-ACGT symbols, $u_t$ is a per-symbol ACGT-vs-unknown gate, and $v_t \in \{0, \ldots, 11\}$ is the unknown-symbol index. This prevents ambiguity codes from contaminating ACGT expert counts while preserving exact 4-bit reversibility.

\subsection{zstandard}

\textit{zstandard}~\citep{zstd} serves as the high-throughput backend for streams where statistical models offer diminishing returns: headers, control bytes, and cases where fast decompression matters more than marginal ratio gains. The codec is applied after 2-bit or 4-bit packing, which gives \hecate an advantage over \textit{NAF}~\citep{kirill2019nucleotide} (which uses exclusively 4-bit packing with a fixed alphabet). Coupled with packing, \textit{zstandard}'s decompression speeds (of order one nanosecond per byte) are not diminished.

\section{Referential Compression}

\hecate supports referential compression of assembled genomes by operating over semantic streams rather than raw FASTA text. Each stream is differenced against the homologous reference stream using \textit{hdiff}~\citep{sisong2013}, which identifies exact-match segments via suffix array lookup with a Bloom filter to skip unnecessary $\mathcal{O}(\log n)$ searches. The output is a series of copy descriptors and unmatched literal segments.

For the nucleotide stream, packed bytes are first expanded to symbol-domain vectors before differencing and re-packed after patching, to preserve exact packed semantics. The unmatched segments are then compressed by any of the codecs above; typically \textit{zstandard} or \textit{hecate-bwt}.

Under a substitution-dominant approximation with mismatch rate $p$ over alphabet $\Sigma$, the conditional entropy of target given reference obeys
\[
H(T \mid R) \approx |T|\bigl[h_2(p) + p \log_2(|\Sigma| - 1)\bigr].
\]
This is the information-theoretic reason referential gains increase superlinearly as similarity grows: for highly similar assemblies (e.g., individual human genomes against GRCh38, with $p \approx 0.006$~\citep{auton2015global}), the gap between $H(T)$ and $H(T \mid R)$ is large enough that even non-trivial patch metadata is amortized. When edits cluster into $r$ literal runs, descriptor cost scales as $\mathcal O(r \log n)$ rather than $\mathcal O(e \log n)$, a further structural advantage for assembled genomes where mismatches are sparse and localized.

\section{Experimental Evaluation}

\subsection{Methodology}

All benchmarks were run on a Zen~3 (Ryzen~9 5950X) CPU with 128 gigabytes of 3200 MT/s DDR4 RAM, performed entirely within memory via a RAM disk, on a quiet temperature-controlled dedicated system.
More details can be found in the supplementary material.
\hecate was compiled with \texttt{cargo build --release}.
We report three metrics: compression ratio in bits per byte ($R_{\mathrm{bpb}} = 8S/N$, lower is better), encoding time in nanoseconds per byte ($t_{\mathrm{enc}}$, lower is better), and decoding time ($t_{\mathrm{dec}}$, lower is better), all using wall-clock time averaged over ten runs.
We use wall-clock time deliberately.
Data compression is commonly assumed to be memory-bound, but as a mostly serial workload\footnote{Decoders typically possess only causal (past) context, limiting the degree of parallelism for processing incoming data.} that induces unavoidable branch and cache misses\footnote{Predictable branches and loads in a noiseless coder would imply redundancy in the compressed stream, which is paradoxical.}, it is at least equally reliant on clock speed and pipeline efficiency.
Reporting wall time rewards implementations that make effective use of available hardware, rather than penalizing parallel codecs for synchronization overhead.

\subsection{Reference-Free Benchmark Results}


We evaluate on seven genomes spanning bacterial, through mammalian to large plant scales:
\begin{itemize}
  \item \texttt{GCF\_000008865.2}: \textit{Escherichia coli} O157:H7 str.\ Sakai; 5.6\,Mb.
  \item \texttt{GCA\_004837865.1}: \textit{Musa balbisiana} isolate DH-PKW; 492.8\,Mb.
  \item \texttt{GCA\_021556685.1}: \textit{Rattus norvegicus} s.\ SHRSP/BbbUtx; 2.9\,Gb.
  \item \texttt{GCF\_000001405.40}: \textit{GRCh38.p14}; 3.3\,Gb.
  \item \texttt{GCF\_000006565.2}: \textit{Toxoplasma gondii} TGA4; 63.7\,Mb.
  \item \texttt{GCA\_000404065.3}: \textit{Pinus taeda} v2.0; 22.5\,Gb.
  \item \texttt{GCF\_009914755.1}: T2T-CHM13v2.0; 3.1\,Gb.
\end{itemize}
\textit{E.~coli} serves as a benchmark for small genomes with limited repetitive structure.
\textit{Musa balbisiana} is a moderately sized plant genome with high repeat content.
The three mammalian genomes stress large-block behaviour, memory locality, and high-order context stability.
Finally, \textit{Pinus taeda} is a large plant genome with extreme repeat content, which tests the limits of BWT-based methods and the benefits of large blocks.

We exclude codecs that do not reach 2\,bpb (e.g., \textit{pigz}, \textit{lbzip2}). The readers are referred to the supplementary material for complete results on all codecs and samples. We were unable to include \textit{GeCo3} in our visualisations due to its slow encode/decode reciprocal throughput (on the tested samples) of 1300--2900\,ns/B.
Figure~\ref{fig:benchmark_gca_000404065_3_fna} shows selected representative benchmark results.

Unlike other comparisons, we do not normalise the input data by removing linebreaks, headers and case information: they are preserved as-is as compared to the original NCBI FASTA files.

\begin{figure*}[t]
    \centering
    \begin{tikzpicture}
        \begin{groupplot}[%
            group style={group size=2 by 5, horizontal sep=4em, vertical sep=4em},
            width=0.45\textwidth,
            height=\hadd\textwidth,ymode=log,
        ]
        \nextgroupplot[title={{Encode (\textit{Pinus taeda})}},
            legend to name=leg:gca_000404065_3_fna,
            legend columns=6,
            legend cell align=left,
            legend style={font=\scriptsize},
        ]
            \addplot+[only marks, mark=*, mark options={solid}, color=Emerald] coordinates {
                (1.346291,22.075804)
                (1.374261,23.146431)
                (1.409751,18.701303)
                (1.414418,13.462651)
                (1.421880,11.765610)
                (1.444403,20.009018)
                (1.480923,12.814075)
                (1.555083,11.888368)
                (1.606766,13.310899)
                (1.649698,19.241946)
                (1.690789,20.756196)
                (1.712998,16.906943)
            };
            \addlegendentry{hecate bwt}
            \addplot+[only marks, mark=*, mark options={solid}, color=OliveGreen] coordinates {
                (1.337551,136.744514)
                (1.341944,145.740595)
            };
            \addlegendentry{markov-mix}
            \addplot+[only marks, mark=*, mark options={solid}, color=Maroon] coordinates {
                (1.362724,116.217480)
                (1.365885,78.794761)
                (1.452247,22.159505)
                (1.532535,110.529566)
                (1.605589,53.230960)
                (1.647987,31.881821)
                (1.673115,5.251833)
                (1.763175,2.050126)
                (1.781973,1.264039)
                (1.783964,53.180228)
                (1.828991,1.225385)
                (1.868202,1.200655)
                (1.868429,1.474137)
                (1.879769,8.917689)
                (1.884075,39.031939)
                (1.905880,2.949384)
                (1.959725,1.530287)
                (1.985258,38.028171)
            };
            \addlegendentry{hecate zstd}
            \addplot+[only marks, mark=o, mark options={solid}, line width=0.75pt, color=Black] coordinates {
                (1.328191,332.144416)
                (1.436969,191.054507)
                (1.447814,181.209211)
            };
            \addlegendentry{MFCompress ($t{=}1$)}
            \addplot+[only marks, mark=square, mark options={fill=none}, line width=0.75pt, color=Gray] coordinates {
                (1.462947,35.448175)
                (1.526066,25.894839)
                (1.544114,24.201519)
            };
            \addlegendentry{MFCompress ($t{=}16$)}
            \addplot+[only marks, mark=o, mark options={solid}, line width=0.75pt, color=VioletRed] coordinates {
                (1.360873,652.989516)
                (1.644301,284.136878)
                (1.857477,3.896124)
                (1.867013,121.100692)
                (1.895495,19.545945)
            };
            \addlegendentry{NAF}

        \nextgroupplot[title={{Decode (\textit{Pinus taeda})}}]
            \addplot+[only marks, mark=*, mark options={solid}, color=Emerald] coordinates {
                (1.346291,7.642660)
                (1.374261,8.032078)
                (1.409751,7.113636)
                (1.414418,7.342989)
                (1.421880,4.640790)
                (1.444403,7.652038)
                (1.480923,7.365045)
                (1.555083,5.062307)
                (1.606766,8.359160)
                (1.649698,9.446988)
                (1.690789,10.283730)
                (1.712998,13.615630)
            };
            \addplot+[only marks, mark=*, mark options={solid}, color=OliveGreen] coordinates {
                (1.337551,82.673821)
                (1.341944,82.718704)
            };
            \addplot+[only marks, mark=*, mark options={solid}, color=Maroon] coordinates {
                (1.362724,2.048198)
                (1.365885,1.601493)
                (1.452247,2.447862)
                (1.532535,2.100439)
                (1.605589,2.087903)
                (1.647987,1.644373)
                (1.673115,3.054979)
                (1.763175,3.252831)
                (1.781973,5.204400)
                (1.783964,2.049846)
                (1.828991,3.880488)
                (1.868202,2.594671)
                (1.868429,1.073247)
                (1.879769,1.910014)
                (1.884075,2.172233)
                (1.905880,1.657556)
                (1.959725,1.480938)
                (1.985258,1.968938)
            };
            \addplot+[only marks, mark=o, mark options={solid}, line width=0.75pt, color=Black] coordinates {
                (1.328191,236.074682)
                (1.436969,176.439875)
                (1.447814,165.779582)
            };
            \addplot+[only marks, mark=square, mark options={fill=none}, line width=0.75pt, color=Gray] coordinates {
                (1.462947,26.713169)
                (1.526066,20.871982)
                (1.544114,21.409908)
            };
            \addplot+[only marks, mark=o, mark options={solid}, line width=0.75pt, color=VioletRed] coordinates {
                (1.360873,1.459797)
                (1.644301,1.367810)
                (1.857477,1.721760)
                (1.867013,1.636883)
                (1.895495,2.053047)
            };
        \nextgroupplot[title={{Encode (\textit{Homo sapiens}, GRCh38)}},
            legend to name=leg:gcf_000001405_40_fna,
            legend columns=5,
            legend cell align=left,
            legend style={font=\scriptsize},
        ]
            \addplot+[only marks, mark=*, mark options={solid}, color=Emerald] coordinates {
                (1.553488,21.939173)
                (1.572002,19.261494)
                (1.596243,23.642110)
                (1.601540,18.886164)
                (1.609916,11.831839)
                (1.609916,11.947525)
                (1.612150,21.130546)
                (1.621678,11.197480)
                (1.621678,11.274523)
                (1.631086,19.756670)
                (1.636874,12.273401)
                (1.642368,20.673268)
                (1.667940,13.355523)
                (1.669956,21.694715)
                (1.672486,14.839745)
            };
            \addplot+[only marks, mark=*, mark options={solid}, color=OliveGreen] coordinates {
                (1.578937,225.761646)
                (1.582738,222.259876)
            };
            \addplot+[only marks, mark=*, mark options={solid}, color=Maroon] coordinates {
                (1.682086,55.884134)
                (1.708600,215.049109)
                (1.712352,6.630857)
                (1.723225,224.164724)
                (1.743397,2.724628)
                (1.744044,34.491219)
                (1.753314,1.432382)
                (1.756937,215.996018)
                (1.779641,53.495604)
                (1.780394,1.295657)
                (1.827050,52.129162)
                (1.836662,1.564821)
                (1.860759,1.196337)
                (1.867999,10.172237)
                (1.889430,3.204385)
                (1.905922,38.926288)
                (1.918472,1.668766)
                (1.975205,38.464626)
            };
            \addplot+[only marks, mark=o, mark options={solid}, line width=0.75pt, color=Black] coordinates {
                (1.600407,382.480475)
                (1.641879,226.571590)
                (1.654660,215.537794)
            };
            \addplot+[only marks, mark=square, mark options={fill=none}, line width=0.75pt, color=Gray] coordinates {
                (1.623667,56.669097)
                (1.658038,37.266384)
                (1.682154,35.236285)
            };
            \addplot+[only marks, mark=o, mark options={solid}, line width=0.75pt, color=VioletRed] coordinates {
                (1.710191,731.328083)
                (1.750441,472.174607)
                (1.830824,4.220390)
                (1.862666,195.937461)
                (1.885392,34.243693)
            };

        \nextgroupplot[title={{Decode (\textit{Homo sapiens}, GRCh38)}}]
            \addplot+[only marks, mark=*, mark options={solid}, color=Emerald] coordinates {
                (1.553488,8.745320)
                (1.572002,8.134819)
                (1.596243,9.241747)
                (1.601540,9.604046)
                (1.609916,5.175370)
                (1.609916,5.192412)
                (1.612150,8.549535)
                (1.621678,5.648324)
                (1.621678,5.649059)
                (1.631086,12.471834)
                (1.636874,5.595892)
                (1.642368,10.412055)
                (1.667940,8.235428)
                (1.669956,12.908200)
                (1.672486,9.406723)
            };
            \addplot+[only marks, mark=*, mark options={solid}, color=OliveGreen] coordinates {
                (1.578937,132.549276)
                (1.582738,133.311311)
            };
            \addplot+[only marks, mark=*, mark options={solid}, color=Maroon] coordinates {
                (1.682086,2.044515)
                (1.708600,1.596737)
                (1.712352,1.991248)
                (1.723225,1.927945)
                (1.743397,1.927382)
                (1.744044,1.425876)
                (1.753314,1.922667)
                (1.756937,1.981681)
                (1.779641,1.733556)
                (1.780394,1.938021)
                (1.827050,1.762065)
                (1.836662,1.197839)
                (1.860759,1.872879)
                (1.867999,1.649896)
                (1.889430,1.640579)
                (1.905922,1.821952)
                (1.918472,1.541479)
                (1.975205,1.751949)
            };
            \addplot+[only marks, mark=o, mark options={solid}, line width=0.75pt, color=Black] coordinates {
                (1.600407,253.925523)
                (1.641879,187.767661)
                (1.654660,182.415446)
            };
            \addplot+[only marks, mark=square, mark options={fill=none}, line width=0.75pt, color=Gray] coordinates {
                (1.623667,42.256295)
                (1.658038,31.126683)
                (1.682154,30.562267)
            };
            \addplot+[only marks, mark=o, mark options={solid}, line width=0.75pt, color=VioletRed] coordinates {
                (1.710191,1.551019)
                (1.750441,1.489637)
                (1.830824,1.628870)
                (1.862666,1.793553)
                (1.885392,3.052634)
            };
        \nextgroupplot[title={{Encode (\textit{Musa balbisiana})}},
            legend to name=leg:gca_004837865_1_fna,
            legend columns=5,
            legend cell align=left,
            legend style={font=\scriptsize},
        ]
            \addplot+[only marks, mark=*, mark options={solid}, color=Emerald] coordinates {
                (1.503511,18.698674)
                (1.503511,18.777618)
                (1.503511,18.865703)
                (1.544746,20.286881)
                (1.544746,20.339782)
                (1.544746,20.446396)
                (1.571769,12.275609)
                (1.571769,12.308033)
                (1.571769,12.333293)
                (1.575226,21.109033)
                (1.596843,11.029243)
                (1.596843,11.031160)
                (1.596843,11.100061)
                (1.601816,11.256231)
                (1.607108,12.881191)
                (1.624737,22.628335)
            };
            \addplot+[only marks, mark=*, mark options={solid}, color=OliveGreen] coordinates {
                (1.473189,251.846448)
                (1.473689,253.178651)
            };
            \addplot+[only marks, mark=*, mark options={solid}, color=Maroon] coordinates {
                (1.500162,874.433821)
                (1.504878,438.529076)
                (1.550949,74.838987)
                (1.559910,72.054377)
                (1.568138,828.270678)
                (1.579430,59.729207)
                (1.609152,17.733914)
                (1.657573,67.027594)
                (1.685445,4.890404)
                (1.701297,1.568459)
                (1.759051,13.626394)
                (1.770214,1.207393)
                (1.776578,40.361924)
                (1.784612,3.124409)
                (1.838112,38.283945)
                (1.847924,1.701530)
                (1.863745,2.005091)
                (1.915174,1.013410)
            };
            \addplot+[only marks, mark=o, mark options={solid}, line width=0.75pt, color=Black] coordinates {
                (1.467571,420.274957)
                (1.515689,233.355872)
                (1.529612,299.004427)
            };
            \addplot+[only marks, mark=square, mark options={fill=none}, line width=0.75pt, color=Gray] coordinates {
                (1.467571,482.077152)
                (1.515689,220.875234)
                (1.529612,306.106170)
            };
            \addplot+[only marks, mark=o, mark options={solid}, line width=0.75pt, color=VioletRed] coordinates {
                (1.503162,1070.660421)
                (1.572263,672.341770)
                (1.751909,271.969723)
                (1.779640,44.784054)
                (1.841473,4.347188)
            };

        \nextgroupplot[title={{Decode (\textit{Musa balbisiana})}}]
            \addplot+[only marks, mark=*, mark options={solid}, color=Emerald] coordinates {
                (1.503511,7.647380)
                (1.503511,7.658146)
                (1.503511,7.678511)
                (1.544746,8.309982)
                (1.544746,8.362772)
                (1.544746,8.396730)
                (1.571769,5.482978)
                (1.571769,5.570827)
                (1.571769,5.634316)
                (1.575226,13.494205)
                (1.596843,5.626294)
                (1.596843,5.646422)
                (1.596843,5.646803)
                (1.601816,5.870227)
                (1.607108,7.582121)
                (1.624737,13.947149)
            };
            \addplot+[only marks, mark=*, mark options={solid}, color=OliveGreen] coordinates {
                (1.473189,159.063023)
                (1.473689,159.764183)
            };
            \addplot+[only marks, mark=*, mark options={solid}, color=Maroon] coordinates {
                (1.500162,1.972639)
                (1.504878,1.631828)
                (1.550949,1.302359)
                (1.559910,1.784884)
                (1.568138,2.037634)
                (1.579430,1.476997)
                (1.609152,1.209119)
                (1.657573,1.847506)
                (1.685445,1.184408)
                (1.701297,1.136189)
                (1.759051,1.756482)
                (1.770214,1.145892)
                (1.776578,1.873590)
                (1.784612,1.637247)
                (1.838112,1.831675)
                (1.847924,1.257987)
                (1.863745,1.585030)
                (1.915174,1.100724)
            };
            \addplot+[only marks, mark=o, mark options={solid}, line width=0.75pt, color=Black] coordinates {
                (1.467571,322.984079)
                (1.515689,207.963885)
                (1.529612,266.831226)
            };
            \addplot+[only marks, mark=square, mark options={fill=none}, line width=0.75pt, color=Gray] coordinates {
                (1.467571,293.655582)
                (1.515689,201.383003)
                (1.529612,279.532228)
            };
            \addplot+[only marks, mark=o, mark options={solid}, line width=0.75pt, color=VioletRed] coordinates {
                (1.503162,2.043379)
                (1.572263,1.582617)
                (1.751909,4.106791)
                (1.779640,3.666062)
                (1.841473,1.802981)
            };
        \nextgroupplot[title={{Encode (\textit{Homo sapiens}, T2T)}},
            legend to name=leg:gcf_009914755_1_fna,
            legend columns=5,
            legend cell align=left,
            legend style={font=\scriptsize},
        ]
            \addplot+[only marks, mark=*, mark options={solid}, color=Emerald] coordinates {
                (1.578616,21.891730)
                (1.591522,19.135255)
                (1.621259,19.040895)
                (1.623206,23.857969)
                (1.635336,21.033646)
                (1.635422,12.023269)
                (1.635422,12.101205)
                (1.647058,8.855301)
                (1.647058,8.937936)
                (1.650160,20.212587)
                (1.659192,12.523420)
                (1.659259,8.915396)
                (1.663327,20.681016)
                (1.687881,11.304901)
                (1.688349,13.361451)
                (1.690807,21.816215)
            };
            \addplot+[only marks, mark=*, mark options={solid}, color=OliveGreen] coordinates {
                (1.622370,224.675310)
                (1.622382,225.643233)
            };
            \addplot+[only marks, mark=*, mark options={solid}, color=Maroon] coordinates {
                (1.711653,61.160525)
                (1.733198,6.772108)
                (1.741151,219.296380)
                (1.756782,231.037821)
                (1.759886,2.660316)
                (1.764650,35.523764)
                (1.770021,1.072747)
                (1.791882,224.128934)
                (1.796891,0.931847)
                (1.798408,55.274760)
                (1.849532,53.662808)
                (1.853003,1.601074)
                (1.878408,0.861011)
                (1.887038,10.435661)
                (1.908876,3.115465)
                (1.923102,39.274196)
                (1.935746,1.692500)
                (1.991992,38.953639)
            };
            \addplot+[only marks, mark=o, mark options={solid}, line width=0.75pt, color=Black] coordinates {
                (1.622515,370.376994)
                (1.653154,233.567209)
                (1.668148,210.579440)
            };
            \addplot+[only marks, mark=square, mark options={fill=none}, line width=0.75pt, color=Gray] coordinates {
                (1.640929,72.858763)
                (1.674636,35.428191)
                (1.699035,37.305423)
            };
            \addplot+[only marks, mark=o, mark options={solid}, line width=0.75pt, color=VioletRed] coordinates {
                (1.740794,766.233643)
                (1.771881,488.039016)
                (1.847133,4.174266)
                (1.881864,205.354875)
                (1.905021,36.937897)
            };

        \nextgroupplot[title={{Decode (\textit{Homo sapiens}, T2T)}}]
            \addplot+[only marks, mark=*, mark options={solid}, color=Emerald] coordinates {
                (1.578616,8.856628)
                (1.591522,7.232371)
                (1.621259,9.906475)
                (1.623206,9.458050)
                (1.635336,7.916927)
                (1.635422,5.188222)
                (1.635422,5.188739)
                (1.647058,3.636217)
                (1.647058,3.657083)
                (1.650160,12.732819)
                (1.659192,5.601768)
                (1.659259,3.617797)
                (1.663327,10.719507)
                (1.687881,6.605747)
                (1.688349,8.318186)
                (1.690807,13.377391)
            };
            \addplot+[only marks, mark=*, mark options={solid}, color=OliveGreen] coordinates {
                (1.622370,135.155919)
                (1.622382,137.221666)
            };
            \addplot+[only marks, mark=*, mark options={solid}, color=Maroon] coordinates {
                (1.711653,1.260620)
                (1.733198,1.159445)
                (1.741151,1.628739)
                (1.756782,1.935460)
                (1.759886,1.130992)
                (1.764650,1.461072)
                (1.770021,1.090452)
                (1.791882,1.947415)
                (1.796891,1.070123)
                (1.798408,1.776949)
                (1.849532,1.739653)
                (1.853003,1.223568)
                (1.878408,0.997221)
                (1.887038,1.695261)
                (1.908876,1.644871)
                (1.923102,1.842390)
                (1.935746,1.574079)
                (1.991992,1.749041)
            };
            \addplot+[only marks, mark=o, mark options={solid}, line width=0.75pt, color=Black] coordinates {
                (1.622515,242.779109)
                (1.653154,182.997601)
                (1.668148,177.684323)
            };
            \addplot+[only marks, mark=square, mark options={fill=none}, line width=0.75pt, color=Gray] coordinates {
                (1.640929,54.346250)
                (1.674636,31.228579)
                (1.699035,27.857483)
            };
            \addplot+[only marks, mark=o, mark options={solid}, line width=0.75pt, color=VioletRed] coordinates {
                (1.740794,1.603172)
                (1.771881,1.576241)
                (1.847133,1.633271)
                (1.881864,1.799608)
                (1.905021,3.139808)
            };
        \nextgroupplot[title={{Encode (\textit{Rattus norvegicus})}},
            legend to name=leg:gca_021556685_1_fna,
            legend columns=5,
            legend cell align=left,
            legend style={font=\scriptsize},
        ]
            \addplot+[only marks, mark=*, mark options={solid}, color=Emerald] coordinates {
                (1.586815,22.267817)
                (1.602948,19.517890)
                (1.627484,23.994126)
                (1.628000,19.110293)
                (1.640625,21.127676)
                (1.642822,11.957324)
                (1.642822,12.294708)
                (1.655807,20.461371)
                (1.657225,9.075805)
                (1.657225,9.133300)
                (1.661266,20.710059)
                (1.666744,12.317831)
                (1.670823,9.118718)
                (1.691491,22.004410)
                (1.693503,13.459731)
                (1.694194,11.167649)
            };
            \addplot+[only marks, mark=*, mark options={solid}, color=OliveGreen] coordinates {
                (1.600534,232.918908)
                (1.600617,234.126393)
            };
            \addplot+[only marks, mark=*, mark options={solid}, color=Maroon] coordinates {
                (1.679230,63.246646)
                (1.690371,233.946009)
                (1.697239,251.553215)
                (1.703584,6.749821)
                (1.714605,36.072187)
                (1.734068,2.664871)
                (1.737608,243.506934)
                (1.742045,1.093054)
                (1.744448,56.001397)
                (1.772759,0.971997)
                (1.802331,54.372725)
                (1.848874,1.599982)
                (1.849884,10.105177)
                (1.866495,3.331393)
                (1.882066,37.904057)
                (1.888097,0.901255)
                (1.896926,1.719714)
                (1.949132,38.448858)
            };
            \addplot+[only marks, mark=o, mark options={solid}, line width=0.75pt, color=Black] coordinates {
                (1.601979,387.876071)
                (1.635089,253.227072)
                (1.648638,221.559040)
            };
            \addplot+[only marks, mark=square, mark options={fill=none}, line width=0.75pt, color=Gray] coordinates {
                (1.615657,63.825171)
                (1.642533,54.681573)
                (1.660909,42.771458)
            };
            \addplot+[only marks, mark=o, mark options={solid}, line width=0.75pt, color=VioletRed] coordinates {
                (1.696844,783.115932)
                (1.717101,497.548305)
                (1.842422,4.120054)
                (1.845059,204.798606)
                (1.862696,36.480988)
            };

        \nextgroupplot[title={{Decode (\textit{Rattus norvegicus})}}]
            \addplot+[only marks, mark=*, mark options={solid}, color=Emerald] coordinates {
                (1.586815,9.199926)
                (1.602948,7.740371)
                (1.627484,9.760659)
                (1.628000,9.533405)
                (1.640625,8.200045)
                (1.642822,5.370635)
                (1.642822,5.382088)
                (1.655807,12.669578)
                (1.657225,3.900606)
                (1.657225,3.927706)
                (1.661266,10.121220)
                (1.666744,6.221816)
                (1.670823,3.845796)
                (1.691491,13.226057)
                (1.693503,8.148899)
                (1.694194,5.945022)
            };
            \addplot+[only marks, mark=*, mark options={solid}, color=OliveGreen] coordinates {
                (1.600534,141.713890)
                (1.600617,144.955927)
            };
            \addplot+[only marks, mark=*, mark options={solid}, color=Maroon] coordinates {
                (1.679230,1.270037)
                (1.690371,1.671581)
                (1.697239,2.018831)
                (1.703584,1.153067)
                (1.714605,1.432917)
                (1.734068,1.131458)
                (1.737608,1.970364)
                (1.742045,1.110784)
                (1.744448,1.748841)
                (1.772759,1.108644)
                (1.802331,1.761219)
                (1.848874,1.229494)
                (1.849884,1.703498)
                (1.866495,1.659657)
                (1.882066,1.807851)
                (1.888097,1.059763)
                (1.896926,1.566354)
                (1.949132,1.776353)
            };
            \addplot+[only marks, mark=o, mark options={solid}, line width=0.75pt, color=Black] coordinates {
                (1.601979,263.365869)
                (1.635089,195.539524)
                (1.648638,190.290107)
            };
            \addplot+[only marks, mark=square, mark options={fill=none}, line width=0.75pt, color=Gray] coordinates {
                (1.615657,47.261739)
                (1.642533,36.912354)
                (1.660909,38.250607)
            };
            \addplot+[only marks, mark=o, mark options={solid}, line width=0.75pt, color=VioletRed] coordinates {
                (1.696844,1.625264)
                (1.717101,1.581109)
                (1.842422,1.733955)
                (1.845059,2.306279)
                (1.862696,2.880301)
            };
        \end{groupplot}
    \end{tikzpicture}\\[0.5em]
    \ref{leg:gca_000404065_3_fna}
    \caption{Encode and decode times for \texttt{GCA\_000404065.3.fna} (\textit{Pinus taeda}, 22.5\,Gb), \texttt{GCF\_000001405.40.fna} (\textit{Homo sapiens} (GRCh38.p14), 3.3\,Gb), \texttt{GCA\_004837865.1.fna} (\textit{Musa balbisiana}, 499\,Mb), \texttt{GCF\_009914755.1.fna} (\textit{Homo sapiens} (T2T-CHM13v2.0), 3.2\,Gb) and \texttt{GCA\_021556685.1.fna} (\textit{Rattus norvegicus}, 2.9\,Gb).}
    \label{fig:benchmark_gca_000404065_3_fna}
\end{figure*}

The pattern across the datasets is consistent.
\textit{hecate-bwt} achieves excellent compression ratios (1.58--1.60\,bpb on CHM13; 1.60--1.61 on \textit{R.~norvegicus}) at a highly competitive throughput.
\textit{markov-mix} reaches the best ratios in our comparison (1.56\,bpb on CHM13) but at higher computational cost.
Compared to MFCompress at similar ratio, \textit{markov-mix} is $2{-}4\times$ faster to encode and $2-3\times$ faster to decode.
\textit{hecate-zstd} matches NAF's operating region while benefiting from 2-bit packing.
The overall picture: \hecate's codecs dominate or tighten the Pareto frontier across the ratio/speed plane, with the strongest gains on large genomes where large-block BWT and high-order Markov models have the most room to exploit long-range structure.

\subsection{Referential Benchmark Results}

We consider GRCh38 as the reference genome for human data, testing against two assemblies from the HPRC:
\begin{itemize}
  \item \texttt{GCA\_044167135.1\_HG01167\_hap1\_hprc\_f2}.
  \item \texttt{GCA\_042077855.1\_HG00133\_hap2\_hprc\_f2}.
\end{itemize}
The dataset totals 9.2\,GB. Table~\ref{tab:referential} summarizes the results.

\begin{table}[t]
  \centering
  \small
  \caption{Referential compression of two human assemblies against GRCh38.}
  \begin{tabular}{@{}lccc@{}}
    \toprule
    Method & Compressed size & Encode (wall) & Decode (wall) \\
    \midrule
    \hecate (\texttt{-EMbwt:100}) & 35.6\,MB & 3:19 & 10.9\,s \\
    AGC (v3.2.0) & 828.5\,MB & 1:00 & 6.2\,s \\
    \bottomrule
  \end{tabular}
  \label{tab:referential}
\end{table}

\hecate achieves a $23\times$ better compression ratio than AGC at comparable aggregate CPU time (530\,s user vs.\ 550\,s). AGC is purpose-built for fast queries on large genome collections and is not optimized for the pairwise referential scenario tested here. The comparison is nonetheless informative: it demonstrates that stream-level differencing followed by strong per-stream coding can substantially outperform collection-oriented tools on individual assembly pairs.

\section{Discussion}

Two limitations are explicit in the current design.
First, the strongest \textit{markov-mix} configuration requires approximately 2.45\,GB of model state.
This is the familiar context-capacity trade-off:
Larger context spaces reduce approximation error in heterogeneous regions but increase cache pressure and resident footprint.
A natural next step is hierarchical state compression (e.g., sparse or shared count slabs) that preserves effective context depth while reducing memory.
Second, referential mode currently materializes full stream payloads before patch/diff, rather than operating in a streaming fashion.
A streaming referential pipeline with block-aligned dependency cuts could reduce memory and improve partial-decoding ergonomics, but requires careful dependency scheduling to avoid ratio regressions from over-constrained patch segmentation.
Both are deliberate first-generation trade-offs:
Algorithmic behaviour stays explicit, deterministic, and reproducible.

More broadly, the results support a systems-level claim:
For genomic data, compression quality depends as much on architecture as on estimator sophistication.
Container semantics, per-stream method assignment, and decode/index constraints affect the achievable practical frontier.
The design of \hecate makes these engineering choices explicit and independently tunable, which likely is why improvements persist across genome scales and similarity regimes.

\section{Conclusion}

The \hecate framework demonstrates that practical genomic compression improves when codec theory and systems architecture are co-designed.
Auxiliary-index BWT with custom multiscale mixing, explicit blockwise Markov expert competition, and semantic-stream referential differencing jointly produce a stronger ratio/speed frontier than monolithic pipelines.
In reference-free mode, \hecate is 2 to 10 times faster than prior tools at the same compression ratio.
Under a fixed time budget, it achieves 5\% to 10\% better compression.
In referential mode, streamwise differencing yields 23 times better compression than AGC on individual assembly pairs.

The central design principle is decomposition:
Stream factorization trades monolithic model mismatch for bounded side channels.
Auxiliary-index BWT provides large-block transform gains with controlled inversion metadata.
The \textit{markov-mix} codec approximates per-block oracle expert choice with bounded selector overhead.
Our referential mode targets $H(T \mid R)$-regime operating points on homologous collections.
This decomposition links algorithmic choices to concrete systems constraints; this is why the improvements we achieve are not attributable to any single trick, but to their coordination.

The \hecate framework is open-source, released under the terms of the Affero General Public License v3.0.
The source code, documentation, and issue tracker can be found at \url{https://gitlab.com/rahmannlab/hecate}.
